\documentclass[
aip,
jap,
amsmath,
amsfonts,
reprint
]{revtex4-1}

\usepackage[english]{babel}
\usepackage{graphicx}% Include figure files
\usepackage{bm}% bold math
\usepackage[colorlinks=true, pdfstartview=FitV, linkcolor=blue,
citecolor=blue, urlcolor=blue]{hyperref}

\begin{document}

%\preprint{}

\title{Boundary between the thermal and statistical polarization
  regimes in a nuclear spin ensemble} %Title of paper



%\author{}
%\thanks{}
%\affiliation{}


\author{B. E. Herzog}
  
\author{D. Cadeddu}

\author{F. Xue}
  
\author{P. Peddibhotla}

\author{M. Poggio}
\email[]{martino.poggio@unibas.ch}
\homepage[]{http://poggiolab.unibas.ch/}

\affiliation{Department of Physics, University of Basel, Klingelbergstrasse 82, 4056 Basel, Switzerland}

\date{\today}

%%%%%%%%%%%%%%%%%%%%%%%%%%%%%%%%%%%%%%%%%%%%%%%%%%%%%%%%%%%%%%%%%%%%%%%%%%%%%%%%%%

\begin{abstract} 
  As the number of spins in an ensemble is reduced, the statistical
  fluctuations in its polarization eventually exceed the mean thermal
  polarization.  This transition has now been surpassed in a number of
  recent nuclear magnetic resonance experiments, which achieve
  nanometer-scale detection volumes.  Here, we measure nanometer-scale
  ensembles of nuclear spins in a KPF$_6$ sample using magnetic
  resonance force microscopy.  In particular, we investigate the
  transition between regimes dominated by thermal and statistical
  nuclear polarization.  The ratio between the two types of
  polarization provides a measure of the number of spins in the
  detected ensemble.
\end{abstract}

%\pacs{\color{red} ???}% insert suggested PACS numbers in braces on next line

%\keywords{}?

\maketitle %\maketitle must follow title, authors, abstract and \pacs

%%%%%%%%%%%%%%%%%%%%%%%%%%%%%%%%%%%%%%%%%%%%%%%%%%%%%%%%%%%%%%%%%%%%%%%%%%%%%%%%%%

%\section{}
%\label{}
%\subsection{}
%\subsubsection{}

% If in two-column mode, this environment will change to single-column format so that long equations can be displayed. 
%\begin{widetext}
%$$\mbox{put long equation here}$$
%\end{widetext}

% Figures should be put into the text as floats. 
% Use the graphics or graphicx packages (distributed with LaTeX2e).
% Put the label that you will use with \ref{} command in the braces of the \label{} command.
%
% \begin{figure}
% \end{figure}

% \begin{table}
% \end{tabular}
% \end{table}

%%%%%%%%%%%%%%%%%%%%%%%%%%%%%%%%%%%%%%%%%%%%%%%%%%%%%%%%%%%%%%%%%%%%%%%%%%%%%%%%%%

%CONTENT

In recent decades, the drive for technological advancement coupled
with an interest in understanding underlying microscopic interactions
has led to rapid growth in the number of studies related to
nanometer-scale phenomena.  The research area broadly known as
nanoscience and nanotechnology brings together a diverse range of
topics including surface science, semiconductor physics, and molecular
self-assembly.  In many systems, physical phenomena at the
nanometer-scale are strikingly different from their behavior at the
macroscale.  In particular, the reduced dimensionality of
nanometer-scale samples can manifest itself in either thermal or
quantum effects not observed in larger systems.  For example, behavior
ranging from the Brownian motion\cite{Frey2005} to the quantization of conductance\cite{VanWees1988}
emerge as measurement length scales are reduced.

The development of nuclear magnetic resonance (NMR) and magnetic
resonance imaging (MRI) with nanometer-scale resolution has been a
particularly captivating goal in nanoscience, due to its potential
impact.  As the only non-destructive, chemically-selective, and truly
three-dimensional imaging technique, MRI is an indispensable tool in a
broad array of fields including medicine, biology, physics, and
materials science.  Conventional inductively-detected MRI, however, is
limited to a detection volume of a few $\mu$m$^3$.\cite{Ciobanu2002} The
extension of this resolution down to a few nm$^3$ and eventually to
atomic resolution has been a long-standing goal.\cite{Sidles1995} The
capability to image molecules atom-by-atom, thus allowing the mapping
of the three-dimensional atomic structure of unknown macro-molecules
would be revolutionary.  While the latter goal has not yet been
achieved, a few experiments in the last few years have demonstrated
nanometer-scale MRI (nanoMRI).\cite{Degen2009,Mamin2009,Nichol2012}
Two techniques, magnetic resonance force microscopy (MRFM) first, and
nitrogen-vacancy (NV) magnetometry shortly thereafter, have both
detected NMR in nanometer-scale detection volumes.\cite{Mamin2007,Mamin2013,Staudacher2013}  
Although so far only MRFM
techniques have produced 3D images of nuclear spin density, e.g.\
virus particles and hydrocarbon layers,\cite{Degen2009,Mamin2009,Nichol2012} NV magnetometry has achieved a
higher sensitivity\cite{Loretz2014} and initial imaging experiments\cite{Rugar2014,Haberle2014,DeVience2014} have recently been made.
 In addition, NV magnetometry
appears particularly promising given its ability to work under ambient
conditions, while high-sensitivity MRFM must be carried out in high
vacuum and at cryogenic temperatures.

Nanometer-scale spin ensembles differ from larger ensembles in that
random fluctuations in the total polarization -- also known as spin
noise -- exceed the normally dominant mean thermal polarization.  This
characteristic imposes important differences between nanoMRI and
conventional MRI protocols.  In the former technique, statistical
fluctuations are usually measured, whereas in the latter the signal is
based on the thermal
polarization.\cite{Bloch1946,Degen2007,Peddibhotla2013} The thermal
polarization -- also known as Boltzmann polarization -- results from
the alignment of nuclear magnetization under thermal equilibrium along
a magnetic field.  The statistical polarization, on the other hand,
arises from the incomplete cancellation of magnetic moments within the
ensemble.  Here, we study the nuclear polarization of nanometer-scale
volumes using MRFM, focusing on the transition between the regimes in
which thermal and statistical polarization dominate.

A single spin in a magnetic field can be described by the Hamiltonian
$\hat{H} = - \hbar \gamma B \hat{I}_z $, where $\hbar$ is the reduced
Plank constant, $\gamma$ the gyromagnetic ratio, $B$ the total
magnetic field and $\hat{I}_z$ the nuclear spin operator along
$\hat{z}$.  For an ensemble of $N$ non-interacting spins, we calculate
the expectation value for $M_z$, i.e. the mean magnetization along
$\hat{z}$, as well as its standard deviation $\sigma_{M_z}$ using the
partition function and density matrix.\cite{Xue2011c,Schlichter1996}
Considering that the thermal energy even at cryogenic temperatures ($T
\sim 1$ K) and high magnetic fields ($B \sim 10$ T) is much larger
than the nuclear Zeeman splitting, i.e.\ $\hbar \gamma B \ll k_B T$,
we neglect orders of $\frac{\hbar \gamma B}{k_B T}$ beyond the first,
resulting in:
\begin{eqnarray}
  M_z & = & N \frac{I(I+1)}{3} \left(\frac{\hbar\gamma B}{k_BT}\right)
  \hbar\gamma, \label{eq1}\\
  \sigma_{M_z} & = & \sqrt{N \frac{I(I+1)}{3}} \hbar\gamma, \label{eq2}
\end{eqnarray}
where $k_B$ is the Boltzmann constant and $T$ the temperature of the
system.

In order to compare the thermal and the statistical polarization, we
express both as fractions of a fully polarized system $M_\text{100\%}=
N \hbar \gamma I$, resulting in $P_\text{thermal} =
\frac{M_z}{M_\text{100\%}} = \frac{I+1}{3}\frac{\hbar\gamma B}{k_BT}$
and $P_\text{statistical} = \frac{\sigma_{M_z}}{M_\text{100\%}} =
\sqrt{\frac{I+1}{3I}\frac{1}{N}}$.  Note that while $P_\text{thermal}$
is independent of the ensemble size, $P_\text{statistical}$ increases
with decreasing ensemble size.  This implies that for ensembles with
$N < N_c$, where $N_c$ is some critical number of spins reflecting the
border of the two regimes, $P_\text{statistical} > P_\text{thermal}$.
For this ensemble size, the size of the natural spin polarization
fluctuations will begin to exceed the magnitude of the mean
polarization in thermal equilibrium.  This transition typically occurs
on the micro- or nanometer-scale, underpinning the dominant role
statistical fluctuations play in nanometer-scale NMR.  Furthermore, by
measuring both mean thermal magnetization (\ref{eq1}) and the standard
deviation (\ref{eq2}), one can determine the number of spins in the
detected ensemble depending on the ratio of $M_z$ and $\sigma_{M_z}$:
\begin{equation}
  N = \frac{3}{ I(I+1) } \left( \frac{ k_B \, T}{\hbar\gamma
      B } \right)^2 \left(\frac{M_z}{\sigma_{M_z}}\right)^2. \label{eq3}
\end{equation}
Note that for $\frac{M_z}{\sigma_{M_z}} = 1$, the ensemble contains $N
= N_c$ spins. In a material with a nuclear spin density $n a$, where
$n$ is the number density of the nuclear element and $a$ is the
natural abundance of the measured isotope, the corresponding detection
volume is then given by $V = \frac{N}{na}$.

\begin{figure}[htp]
	\includegraphics[width=8.5cm]{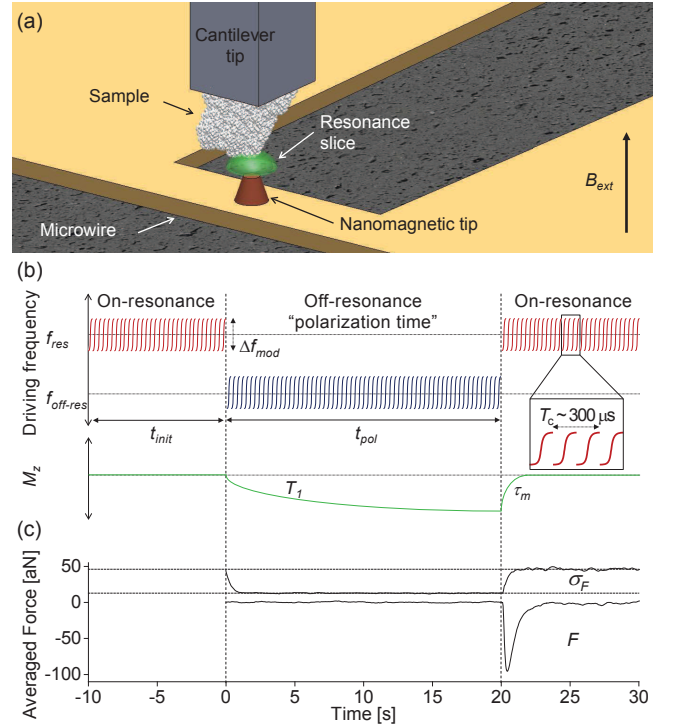}
	\caption{\label{f:expsetup} \textbf{(a)} The experimental
          set-up with the poly-crystalline KPF$_{\text{6}}$ sample
          (white) at the end of the cantilever.  A small section of it
          intersects with the resonance slice (green) above the
          nanomagnet.  The microwire produces the transverse rf
          magnetic field used to adiabatically invert the nuclear
          spins. \textbf{(b)} Schematic diagram of the pulse sequence
          and the response of the average nuclear magnetization $M_z$.
          The pulse spacing has been exaggerated for clarity.
          \textbf{(c)} $F$ and $\sigma_F$ averaged over 500
          measurements for $t_{\text{pol}} = 20$ s and $\Delta f = 3$
          MHz.  The fast decay of $F$ after the pulses are switched on
          resonance is due to the correlation time of the spins during
          the ARP pulses, $\tau_m$.  }
      \end{figure}

      We measure nanometer-scale volumes of $^{19}$F spins in a sample
      of KPF$_6$ by MRFM.  The ($1.2 \times 1.4 \times
      3.2$)-$\mu$m$^3$ sample is glued to the end of an
      ultra-sensitive Si cantilever.  The cantilever is
      130-$\mu$m-long, 4-$\mu$m-wide, 0.1-$\mu$m-thick and has a
      spring constant $k = 75$ $\mu$N/m, as determined by thermal
      noise measurements at various temperatures.  In the cryogenic
      measurement chamber at $T = 4.4$ K and in a vacuum better than
      $10^{-6}$ mbar, the sample-loaded cantilever has a mechanical
      resonance frequency $f_c = 3.28$ kHz and a quality factor $Q =
      3.1 \times 10^4$.  The apparatus includes a fiber-optic
      interferometer to measure the cantilever's displacement $x$ and
      a superconducting magnet for the application of an external
      field up to $B_{\text{ext}} = 6$ T along the cantilever axis
      $\hat{z}$.  Immediately beneath the sample -- at a distance of
      50-100 nm -- a nanomagentic tip integrated on top of a metallic
      microwire produces strong spatial magnetic field gradients, as
      shown in Fig.~\ref{f:expsetup}.\cite{Poggio2007a} To reduce
      electrostatic interactions between the magnetic tip and the
      sample, a 15-nm-thick layer of Au is evaporated on the sample
      after attachment.

      The microwire acts as a radio frequency source for the
      application of adiabatic rapid passage (ARP) pulses of the
      transverse field to the spin
      ensemble.\cite{Kupce1996,Tannus1996} We drive current through
      the microwire with the frequency-sweep waveform shown in
      Fig.~\ref{f:expsetup}(b).  By synchronizing the ARP pulses such
      that they produce a transverse rf magnetic field whose frequency
      is swept through the nuclear magnetic resonance twice every
      cantilever period, $1/f_c = T_c$, we drive longitudinal nuclear
      spin flips in the sample at $f_c$.  Since the sample is affixed
      to the end of the cantilever, in the presence of the large
      magnetic field gradient $\frac{\partial B}{\partial x}$
      generated by the nanomagnetic tip, the spin flips produce a
      alternating force that drives the cantilever's mechanical
      resonance.  By measuring the amplitude of the cantilever's
      resonant oscillations $x(t)$ with the fiber interferometer and a
      lock-in amplifier, we therefore determine the force acting on it
      $F(t) = \frac{k}{Q} x(t)$.  Note that during the measurement we
      damp the cantilever using electronic feedback to a quality
      factor $Q = 400$ in order to increase the bandwidth $\Delta
      f_{\text{meas}}$ of our force detection without sacrificing
      force sensitivity.\cite{Poggio2007a} From $F(t)$ we derive the
      average force $F$ and the its standard deviation $\sigma_F$ over
      a fixed time intervals.

      The volume of spins, which cyclically inverts at $f_c$ due to
      the ARP pulses, is known as the resonant slice.  The position
      and volume of this slice is determined by the spatial dependence
      of the magnetic field produced by the nanomagnetic tip
      $B_{\text{tip}}$ and by the parameters of the pulses.  A
      schematic representation of the resonant slice, the nanomagnetic
      tip, and the ARP pulse sequence are shown in
      Fig.~\ref{f:expsetup}.  The intersection of the resonant slice
      with the sample constitutes the volume of spins addressed by the
      ARP pulses and therefore the NMR detection volume $V$.  Given
      the parameters used in these experiments, $V$ is always
      concentrated to a small region of space less than $(100 \text{
        nm})^3$.

      When the ARP pulses are tuned to the NMR frequency of nuclei
      inside the resonant slice, the mean and standard deviation of
      the force acting on the cantilever, $F$ and $\sigma_{F}$
      respectively, depend on the mean and standard deviation of the
      spin ensemble's magnetization, $M_z$ and $\sigma_{M_z}$: $F =
      \frac{\partial B}{\partial x} M_z$ and $\sigma_{F} =
      \sqrt{\sigma_{\text{spin}}^2 + \sigma_{\text{cant}}^2} = \sqrt{
        \left ( \frac{\partial B}{\partial x} \right )^2
        \sigma_{M_z}^2 + \sigma_{\text{cant}}^2}$, where $B =
      B_{\text{ext}} + B_{\text{tip}}$ is the total magnetic field in
      the detection volume,\cite{EndnoteAveraging} $\hat{x}$ is the
      direction of the cantilever oscillation, and by the
      fluctuation-dissipation theorem $\sigma_{\text{cant}} =
      \sqrt{\frac{2 k k_B T \Delta f_{\text{meas}}}{\pi f_c Q}}$ is
      the standard deviation of the random thermal force acting on the
      cantilever (the lock-in bandwidth $\Delta f_{\text{meas}} = 0.1$
      Hz must fulfill $\Delta f_{\text{meas}} < f_c/Q$ for the damped
      $Q = 400$).  Off resonance, only thermal fluctuations drive the
      cantilever resulting in $F = 0$ and $\sigma_F =
      \sigma_{\text{cant}}$.

\begin{figure}[htp]
	\includegraphics[width=8.5cm]{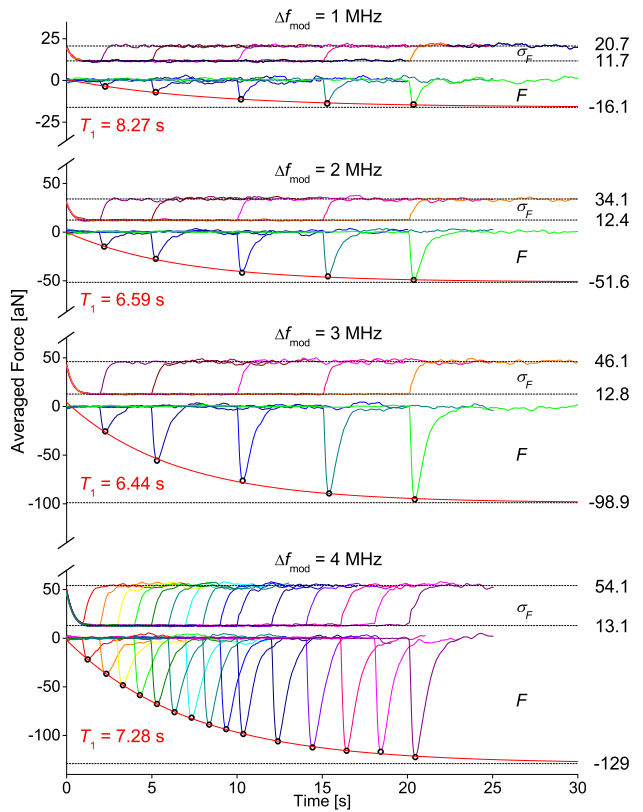}
	\caption{\label{f:TP1-4} TEST $F$ and $\sigma_F$ averaged over 500
          measurements at $B = 4.37$ T and $T = 4.4$ K for different
          modulation widths $\Delta f_\text{mod} = \{1,2,3,4\}$ MHz of
          the ARP pulses.  Each graph shows $F$ in the lower part,
          reflecting the thermal polarization, and $\sigma_F$ in the
          upper part, reflecting the statistical polarization, for a
          series of different polarization times $t_\text{pol}$.  Only
          at $t = t_\text{pol}$ (black circles), when the ARP pulses
          are turned back on resonance, are $F$ and $\sigma_F$ related
          to the thermal and statistical spin polarization
          respectively.  During the polarization time ($t <
          t_{\text{pol}}$) $\sigma_F = \sigma_\text{cant}$, while on
          resonance ($t > t_{\text{pol}}$) spin noise also
          contributes, i.e.\ $\sigma_F = \sqrt{\sigma_{\text{spin}}^2
            + \sigma_{\text{cant}}^2}$.}
\end{figure}

In order to measure the size of the thermal nuclear polarization, we
first initialize the spins to a mean polarization of zero by applying
the ARP pulse sequence with its carrier frequency set to
$f_{\text{res}}$, the NMR frequency of the nuclear spins of interest.
During the application of the resonant ARP pulses, the nuclear spins
have a short correlation time $\tau_m \approx 200$ ms.  Therefore, by
applying resonant pulses for a time $t_{\text{init}} \gg \tau_m$, the
initial thermal polarization is erased, leaving only the statistical
polarization fluctuations.  $\tau_m \ll T_1$ and is limited by the
relaxation time in the rotating frame $T_{1 \rho}$ and the
adiabaticity of the ARP pulses.\cite{Degen2008} At $t=0$ we change the
carrier frequency to $f_{\text{off-res}}$ far from $f_{\text{res}}$,
as shown in Fig.~\ref{f:expsetup}(b).  During this off-resonant time
$t_\text{pol}$, the spin ensemble polarizes along the magnetic field
with a characteristic time $T_1$.  By allowing the off-resonant
condition to persist for a variety of different $t_{\text{pol}}$
before tuning the ARP pulses back on-resonance and measuring the
resulting $F$, we can measure the build-up of the ensemble's thermal
polarization.  In Fig.~\ref{f:TP1-4}, we show measurements of
nanometer-scale ensembles of $^{19}$F nuclear spins at $B = 4.37$ T
and $T = 4.4$ K.  The on- and off-resonant carrier frequencies are
$f_{\text{res}} = 175$ MHz and $f_{\text{off-res}} = 168$ MHz
respectively.  Different ensemble sizes are addressed by changing the
frequency modulation amplitude $\Delta f_{\text{mod}}$ of the ARP
pulse sequences.  The thickness of the resonant slice and therefore
its volume of intersection with the sample is roughly proportional to
$\Delta f_{\text{mod}}$.  From fits to these signals, shown in
Fig. \ref{f:TP1-4}, we can extract both the force $F$ due to the
equilibrium thermal polarization and the spin-lattice relaxation time
$T_1$.  By plotting the standard deviation of the resonant force
$\sigma_F$ during the same experiments, we can also measure the effect
of the ensemble's statistical fluctuations $\sigma_{M_z}$.  As
expected from (\ref{eq1}) and (\ref{eq2}), for detection volumes with
nearly constant $\frac{\partial B}{\partial x}$, $F$ increases
linearly with increasing detection volume or roughly linearly with
$\Delta f_{\text{mod}}$, while $\sigma_{\text{spin}}$ increases
roughly as $\sqrt{\Delta f_{\text{mod}}}$.

\begin{figure}[htp]
	\includegraphics[width=8.5cm]{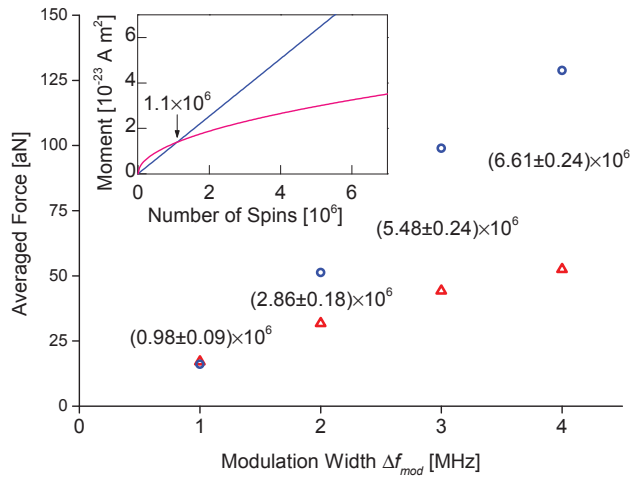}
	\caption{\label{f:mean+std_vs_modwidth} $F$ (blue circles),
          originating from the thermal polarization, and
          $\sigma_{\text{spin}}$ (red triangles), originating from the
          statistical polarization, as a function of the ARP
          modulation width $\Delta f_\text{mod}$ at $B = 4.37$ T and
          $T = 4.4$ K.  The values between the symbols show the
          corresponding number of spins $N$ given by equation
          (\ref{eq3}).  \textbf{Inset:} A theoretical plot of $M_z$
          and $\sigma_{M_z}$ for $^{19}$F as a function of $N$ showing
          the crossover at $N_c = 1.10 \times 10^6$ spins.  The
          similarity between the inset and the figure indicate that
          the number of detected spins or the detection volume $V$ is
          roughly proportional to $\Delta f_{\text{mod}}$.}
\end{figure}

From the ratio of the measured thermal and statistical polarizations
and using equation (\ref{eq3}) with $\frac{M_z}{\sigma_{M_z}} =
\frac{F}{\sigma_{\text{spin}}}$, we determine the number of spins in
the detected ensembles. $N$ ranges from $0.98 \times 10^6$ to $6.61
\times 10^6$ corresponding to detection volumes $V$ from $(26.3 \text{
  nm})^3$ to $(49.7 \text{ nm})^3$.  As shown in
Fig.~\ref{f:mean+std_vs_modwidth}, pulses with the smallest $\Delta
f_{\text{mod}} = 1$ MHz address a spin ensemble slightly smaller than
$N_c = 1.10 \times 10^6$ spins, i.e. just small enough to be dominated
by statistical nuclear spin polarization.  The calculated number of
spins compare favorably to the lower limit of spins determined through
estimates of the magnetic field gradient based on a magnetostatic
model in the manner of the supplementary section of Peddibhotla
\textit{et al}.\cite{Peddibhotla2013}.

Experiments also show that $T_1 = 7.2 \pm 1.0$ s and is independent
of the ensemble size within the error of the measurement.  This value
is similar to previous measurements of larger detection volumes,
yielding $T_1 = 6.2$ s.\cite{Eberhardt2007}
Precise comparisons are difficult given that $T_1$ depends strongly on
the density of paramagnetic impurities in the sample as well as oxygen
at the surface.  Recent measurements of small ensembles of electron
spins, also show that a small detection volume can alter the
measured $T_1$ relative to conventional measurements.  For tiny
detection volumes within a larger sample, the measured $T_1$ can be
reduced by spin diffusion effects relative to measurements of
macroscopic detection volumes.\cite{Cardellino2014}

For detection volumes in which $\frac{\partial B}{\partial x}$ is
nearly constant, the error in the determination of the size of the
detected ensemble depends only on the error of the measurements of $F$
and $\sigma_F$ and on the error in determining $B$ and $T$.  This
method therefore provides a complementary and, in some cases, more
precise alternative to other techniques.  In particular, in MRFM the
size of the detected ensemble is usually determined by measuring $F$
or $\sigma_F$ (depending on whether the volume is in the thermal or
statistical regime), estimating the magnetic field gradient, and
calculating the number of moments responsible for the measured force.
The precision of this scheme depends on knowledge of the magnetic
field gradient at the sample and the spring constant of the
cantilever.  Often, such quantities are measured with a high degree of
error.  An estimate of the size of the detection volume can also be
made through knowledge of the magnetic field profile of the tip,
calculation of the resonant slice geometry, and knowledge of the shape
and position of the sample.  Again, such calculations are typically
imprecise.  In fact, our method can be applied to any NMR technique
capable of detecting both the thermal and statistical polarizations.
These include conventional RF probes at room temperature
\cite{McCoy1989,Gueron1989} and optical Faraday rotation methods in
alkali metal vapors.\cite{Crooker2004} In all cases, the comparison of
statistical and thermal magnetization may provide additional
information, especially when either the precise shape or density
distribution of the sample is not known.

In conclusion, we perform NMR measurements of small ensembles of
$^{19}$F nuclei showing the transition from a thermally dominated to
statistically dominated ensemble magnetization.  In addition, we
demonstrate a method for determining the number of spins in
nanometer-scale ensembles by measuring and comparing both the thermal
and statistical polarizations.  These results are relevant to a number
of recent experiments, which can now address nanometer-scale ensembles
of nuclear spins.  Until today, statistical polarization in
conventional NMR and MRI of macroscopic samples has played a limited
role.\cite{Muller2006} The fact that even for a fairly large ensemble
of $10^6$ $^{19}$F nuclear spins at low temperature and high field
natural polarization fluctuations overtake the thermal polarization
underscores just how weak conventional NMR signals are.  As methods
for nanoMRI continue to develop, the role of statistical polarization,
as highlighted here, will become increasingly important.

\begin{acknowledgments}
 % The authors thank Dr. Fei Xue for experimental assistance. 
  We acknowledge support from the Sino Swiss Science and Technology
  Cooperation (Project IZLCZ2 138894), the Swiss Nanoscience Institute
  (Project P1207), and the Swiss National Science Foundation (Grant
  No. 200020-140478).
\end{acknowledgments}

%\bibliography{herzogbib,EndnoteStatVsBoltz2014}

\begin{thebibliography}{30}
\bibitem [{End()}]{EndnoteAveraging}%
  \BibitemOpen
  \href@noop {} {\bibinfo  {journal} {{In general $F = \left \langle
  \frac{\partial B}{\partial x} \right \rangle_V M_z$ and $\sigma_{F} = \sqrt{
  \left \langle \left ( \frac{\partial B}{\partial x} \right )^2 \right
  \rangle_V \sigma_{M_z}^2 + \sigma_{\text{cant}}^2}$, where $\left \langle
  \cdot \cdot \cdot \right \rangle_V$ denotes an average taken over the
  detection volume $V$. For detection volumes small enough that $\frac{\partial
  B}{\partial x}$ is slowly varying throughout, as in all experiments discussed
  here, $\left ( \left \langle \frac{\partial B}{\partial x} \right \rangle_V
  \right )^2 = \left \langle \left (\frac{\partial B}{\partial x} \right)^2
  \right \rangle_V$}}\ }\BibitemShut {NoStop}%
\end{thebibliography}

%

\end{document}